\documentclass[prl,preprintnumbers,twocolumn,eqsecnum,floatfix,a4paper,nofootinbib,superscriptaddress]{revtex4}
\usepackage{color}
\usepackage{calc}
\usepackage{amsmath,amssymb,graphicx}
\usepackage{amssymb,amsmath}
\usepackage{bm}
\usepackage{microtype}
\usepackage{booktabs}
\usepackage{times}
\usepackage[varg]{txfonts}
\usepackage[colorlinks, pdfborder={0 0 0}]{hyperref}
\usepackage[utf8]{inputenc}
\definecolor{LinkColor}{rgb}{0.75, 0, 0}
\definecolor{CiteColor}{rgb}{0, 0.5, 0.5}
\definecolor{UrlColor}{rgb}{0, 0, 0.75}
\hypersetup{linkcolor=LinkColor}
\hypersetup{citecolor=CiteColor}
\hypersetup{urlcolor=UrlColor}
\allowdisplaybreaks
\DeclareFontFamily{OT1}{pzc}{}
\DeclareFontShape{OT1}{pzc}{m}{it}{<-> s * [1.10] pzcmi7t}{}
\DeclareMathAlphabet{\mathpzc}{OT1}{pzc}{m}{it}

\newcommand{\h}{\mathpzc{h}}

\newcommand{\hlm}{\mathpzc{h}_{\ell m}}

\newcommand{\Ylm}{{Y}^{-2}_{\ell m}}

\newcommand{\blambda}{\bm{\lambda}}
\newcommand{\btheta}{\bm{\theta}}
\newcommand{\bxi}{\bm{\xi}}
\newcommand{\etal}{\emph{et al}}
\newcommand{\n}{\mathbf{n}}

\begin{document}

\title{A ``no-hair'' test for binary black holes}

\author{Siddharth Dhanpal}
\affiliation{International Centre for Theoretical Sciences, Tata Institute of Fundamental Research, Bangalore 560012, India}
\affiliation{UM-DAE Centre for Excellence in Basic Sciences, Mumbai 400098, India}
\author{Abhirup Ghosh}
\author{Ajit Kumar Mehta}
\affiliation{International Centre for Theoretical Sciences, Tata Institute of Fundamental Research, Bangalore 560012, India}
\author{Parameswaran~Ajith}
\affiliation{International Centre for Theoretical Sciences, Tata Institute of Fundamental Research, Bangalore 560012, India}
\affiliation{Canadian Institute for Advanced Research, CIFAR Azrieli Global Scholar, MaRS Centre, West Tower, 661 University Ave., Suite 505, Toronto, ON M5G 1M1, Canada}
\author{B.~S.~Sathyaprakash}
\affiliation{Department of Physics and Department of Astronomy and Astrophysics, The Pennsylvania State University, University Park, PA 16802, USA}
\affiliation{School of Physics and Astronomy, Cardiff University, Cardiff, CF24 3AA, UK}

\begin{abstract}
One of the consequences of the black-hole ``no-hair'' theorem in general relativity (GR) is that gravitational radiation (\emph{quasi-normal modes}) from a perturbed Kerr black hole is uniquely determined by its mass and spin. Thus, the spectrum of quasi-normal mode frequencies have to be all consistent with the same value of the mass and spin. Similarly, the gravitational radiation from a coalescing binary black hole system is uniquely determined by a small number of parameters (masses and spins of the black holes and orbital parameters). Thus, consistency between different spherical harmonic modes of the radiation is a powerful test that the observed system is a binary black hole predicted by GR. We formulate such a test, develop a Bayesian implementation, demonstrate its performance on simulated data and investigate the possibility of performing such a test using previous and upcoming gravitational wave observations. 
\end{abstract}
\preprint{LIGO-P1800056-v4}
\maketitle
%%%%%%%%%%%%%%%%%%%%%%%%%%%%%%%%%%%%%%%%%%%%%%%%%%%%%%%%%%%%%%%%%%%%%%%%%%%%%%%%%%%%%%%%%%%%%%%%%%%%%%%%%%%%%%%%%%%%%%%%%%%%%%%%%%%%%%%%%%%%%%%`
\paragraph{Introduction:---}

One of the remarkable predictions of general relativity (GR) is that a stationary black hole can be fully described by a small number of parameters --- its mass, spin angular momentum and electric charge~\cite{Israel:1967,Israel:1968,Carter:1978}. As a consequence of this ``no-hair'' theorem, frequencies of the gravitational radiation (\emph{quasi-normal modes}~\cite{Vishveshwara:1970zz,Press:1971wr,Chandrasekhar:1975zza}) from a perturbed black hole is fully determined by these parameters. Astrophysical black holes are not expected to possess significant electric charge; hence, different quasi-normal modes have to be consistent with the same value of the mass and spin. Thus, the consistency between multiple quasi-normal modes provides a test of the ``no-hair'' theorem for stationary, isolated black holes~\cite{Dreyer:2003bv}. Similarly, the dynamics and gravitational radiation from a binary black hole system are uniquely determined by a small number of parameters (masses and spins of the black holes and orbital parameters), and hence different spherical harmonic modes of the radiation have to be consistent with the same values of this small set of parameters. Thus, the consistency between different modes of the observed signal is a powerful test that the radiation emanated from a binary black hole. Inconsistency between different modes would point to either a departure from GR, or the non-black hole nature of the compact objects. 

Coalescence of binaries composed of chargeless black holes would produce a perturbed Kerr black hole as the remnant, and the late time gravitational-wave (GW) signal is described by a spectrum of quasi-normal modes (see, e.g.~\cite{Buonanno:2006ui}). While the relatively simple structure of quasi-normal modes has been known from black-hole perturbation theory for a long time~(see, e.g., \cite{Berti:2009kk} for a review), the radiation from the full coalescence (inspiral, merger and ringdown) have a much more complex structure. Fortunately, recent numerical-relativity simulations, together with high-order analytical calculations, have produced semi-analytical waveforms describing the many of the subdominant multipoles of the radiation that are relevant for observations~\cite{Pan:2011gk,London:2017bcn,Mehta:2017jpq}. The availability of such waveforms allows a powerful test of GR based on the consistency of different modes of the radiation.  

\paragraph{Testing the consistency between different multipoles of the gravitational radiation:--}

%%%%%%%%%%%%%%%%%%%%%%%%%%%%%%%%%%%%%%%%%%%%%%%%%%%%%%%%%%%%%%%%%%%%%%%%%%%%%%%%%%%%%%%%%%
\begin{figure}[htb] \begin{center}
\includegraphics[width=3.4in]{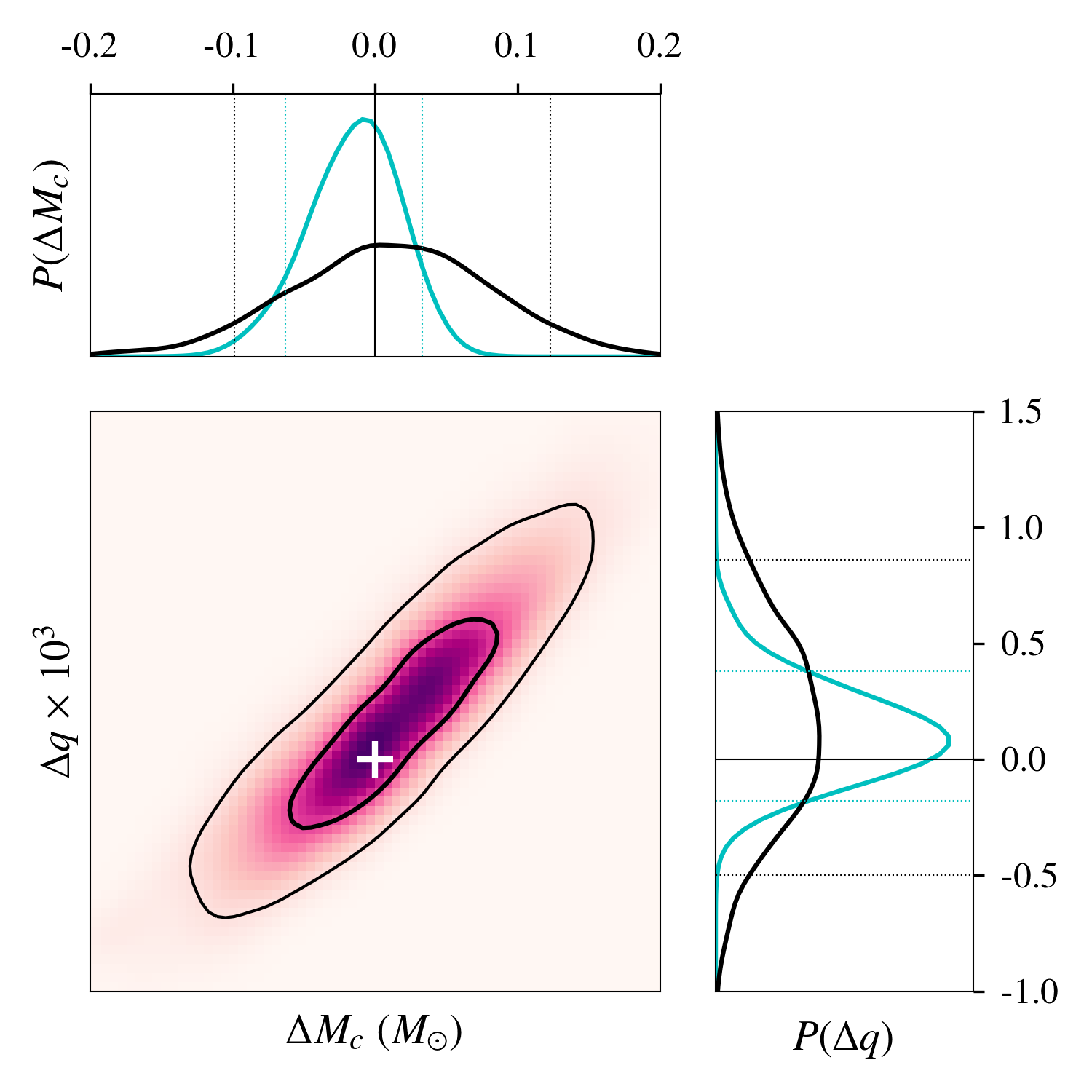}
\caption{In the middle panel, the thick (thin) contours show the 50\% (90\%) credible regions in the joint posteriors of two parameters $\Delta M_c$ and $\Delta q$ that describe discrepancies in the estimated parameters using the quadrupole and non-quadrupole modes, estimated from a simulated GR signal. Black histograms on the side panels show the marginalized posteriors in $\Delta M_c$ and $\Delta q$, while the cyan histograms show the 1-dimensional posteriors in $\Delta M_c$ and $\Delta q$ estimated from the data by introducing only one variation (say, $\Delta M_c$) at a time, keeping the other fixed (say, $\Delta q = 0$). It can be seen that the posteriors are fully consistent with the GR prediction of $\Delta M_c = \Delta q = 0$ (shown by a ``+'' sign in the center panel and by thin black lines in side panels). In the side panels, the dotted lines mark the 90\% credible regions. The simulated GR signal corresponds to a binary with total mass $M = {80}M_\odot$ and mass ratio $q = 1/9$ and an inclination angle $\iota = {60^\circ}$ observed by a single Advanced LIGO detector with an optimal SNR of 25. }
\label{fig:contour_plots}
\end{center} \end{figure}
%%%%%%%%%%%%%%%%%%%%%%%%%%%%%%%%%%%%%%%%%%%%%%%%%%%%%%%%%%%%%%%%%%%%%%%%%%%%%%%%%%%%%%%%%%
In practice it is very difficult to extract different multipoles of the radiation from the GW observation of a single binary black hole system --- all we measure is a particular linear combination of the modes. Thus, our strategy, developed below, is to introduce extra parameters that describe inconsistency between different modes and to constrain them using a Bayesian framework. This is similar in spirit to the tests of the ``no-hair'' theorem using quasi-normal modes, developed in Refs.~\cite{Gossan:2011ha,Meidam:2014jpa}.

The two polarizations $h_+(t)$ and $h_\times(t)$ of gravitational radiation in GR can be written as a complex time series $\h(t) := h_+(t) - i \, h_\times(t)$, which can be expanded in a basis of spin $-2$ weighted spherical harmonics~\cite{NewmanPenrose} as:
\begin{equation}
\h(t; \n, \blambda) =  \frac{1}{d_L} \sum _{\ell=2}^{\infty} \sum _{m=-\ell}^{\ell} \Ylm (\n) \, {{\hlm}(t; \blambda)}, 
\label{eq:spherical_harmonics}
\end{equation}
where $\Ylm$ are the basis functions of spin $-2$ spherical harmonics, $\n := \{\iota, \varphi_0\}$ define the direction of radiation in the source frame, $d_L$ is  the luminosity distance to the binary, and ${\h}_{lm}(t; \blambda)$ are the spherical harmonic modes of the waveform, which are completely described by the intrinsic parameters $\blambda$ of the system. We assume that the black holes are non-spinning and the binary to be quasi-circular. Hence $\blambda$ consists of only the masses $m_1$ and $m_2$ of the black holes (it is more convenient to describe the system in terms of the \emph{chirp mass} $M_c := {(m_1m_2)^{3/5}}/{(m_1+m_2)^{1/5}}$ and mass ratio $q = m_2/m_1 \leq 1$). In GR, the gravitational radiation is dominated by the quadrupole modes ($\ell = 2, m = \pm 2$); however non-quadrupole modes can make an appreciable contribution if the black holes have significantly unequal masses. The set of intrinsic parameters $\blambda := \{M_c, q\}$ completely determines the multipolar structure (i.e., spherical harmonic modes) of the waveform ${\h}_{lm}(t)$. 

In order to formulate a consistency test between different multipoles, we rewrite Eq.~(\ref{eq:spherical_harmonics}) by splitting the contributions from the dominant $(\ell = 2, m = \pm 2)$ mode of gravitational radiation, and the sub-dominant (higher order) modes 
\begin{eqnarray}
\h(t; \n, \blambda, \Delta \blambda) & = & \sum_{m = \pm2} Y^{-2}_{2m} (\n) {\h}_{2m}(t, \blambda)  \nonumber \\ 
& + & \sum _{\text{H.O.M}} \Ylm (\n) \hlm(t, \blambda+\Delta \blambda)
\label{eq:test_HM}
\end{eqnarray}
where the sum in the second term on the RHS is just over the higher-order modes (H.O.M). Note that we allow a possibility of inconsistency between the dominant mode and higher order modes by introducing a deviation $\Delta \blambda := \{\Delta M_c, \Delta q\}$ in the set of intrinsic parameters that describe the higher order modes; in GR,  $\Delta \blambda = 0$. 

An interferometric GW detector observes a linear combination of the two polarizations $h_+(t)$ and $h_\times(t)$, given by 
\begin{equation}
h(t) = F_+(\theta, \phi, \psi) \, h_+(t-t_0) + F_{\times}(\theta, \phi, \psi)\, {h}_{\times}(t-t_0), 
\label{eq:det_response}
\end{equation}
where $F_+$ and $F_\times$ are the antenna pattern functions of the GW detector, $t_0$ is the time of arrival of the signal at the detector, and $(\theta, \phi), \psi$ define the sky position and polarisation angle of the GW source, respectively. For coalescing binary black hole (BBH) systems in quasi-circular orbits, the observed signal $h(t)$ is described by a set of \emph{intrinsic} parameters $\blambda = \{M_c, q\}$ and \emph{extrinsic} parameters  $\btheta := \{t_0, \iota, \varphi_0, d_L, \theta, \phi, \psi\}$ in GR. 
%If we consider observation using a single detector (as we do in this paper), some of the extrinsic parameters are degenerate with others, and hence it is possible to express the observed signal in terms of a smaller number of effective parameters $\btheta_\mathrm{eff} := \{t_0, \iota, \varphi_0, d_\mathrm{eff}, \psi_\mathrm{eff}\}$, where $d_\mathrm{eff} := d_L \, (F_+^2 + F_\times^2)^{-1/2}$ and $\psi_\mathrm{eff}:= \arctan (F_\times/F_+)$. 
In addition to the parameters that describe signals in GR, we introduce a set of parameters $\Delta \blambda$ describing difference between the intrinsic parameters used to generate the dominant and subdominant modes. The combined set of parameters is denoted as $\bxi = \{\blambda, \btheta, \Delta \blambda\}$. 

The data $d(t) = n(t) + h(t)$ contains the observed signal $h(t)$ given in Eq.~(\ref{eq:det_response}) along with noise $n(t)$, which is modeled as a stationary Gaussian random process.
Given data $d$ and assuming a particular model of the waveform given in \eqref{eq:test_HM} as our hypothesis $H$, we can compute the posterior distribution of the set of parameters ${\bxi}$ making use of the Bayes theorem, which states: 
\begin{equation}
P({\bxi} \, | \, d, H) = \frac{P({\bxi} \, | \, H) \, P (d \, | \, {\bxi}, H)}{P(d \, | \, H)}.
\label{eq:Bayes_theorem}
\end{equation} 
The \emph{posterior} probability density $P({\bxi}\,|\,d,H)$ that the data contains a signal with parameters $\bxi$ is determined by the \emph{prior} probability distribution $P({\bxi} \, | \, H)$ and the \emph{likelihood} $P (d \, | \, {\bxi}, H)$ that the data contains a signal described by parameters $\bxi$; $P(d \, | \, H)$ is a normalization constant, called the \emph{evidence}. For stationary Gaussian noise with power spectral density $S_n(f)$, the likelihood can be written as:
\begin{equation}
P (d \, | \, {\bxi}, H) = \text{exp}\Big[ -\frac{1}{2}\int_{f_\mathrm{low}}^{f_\mathrm{high}} \frac{|\tilde{d}(f) - \tilde{h}(f;{\bxi}, H)|^2}{S_n(f)}df\Big]
\end{equation}
where $f_\mathrm{low}$ and $f_\mathrm{high}$ define the sensitivity bandwidth of the detector, while $\tilde{d}(f)$ and $\tilde{h}(f)$ are the Fourier transforms of $d(t)$ and $h(t)$, respectively. 

Using the above definition for the likelihood function, one proceeds to estimate $\bxi$ by stochastically sampling over the entire parameter space of interest. In this work, we use the \texttt{emcee}~\cite{foreman2013emcee} package, a Python implementation of the affine-invariant ensemble sampler for Markov chain Monte Carlo (MCMC) proposed by \cite{goodman2010ensemble}. This code can be easily parallelized to use multiple computing cores, giving it a major advantage over traditional MCMC algorithms~\footnote{We have compared the posterior distributions obtained from our \texttt{emcee} based code with that from the Nested-Sampling based \textsc{LALInferenceNest} code~\cite{Veitch:2009hd} that is part of the LIGO Algorithm Library (LAL) software suite~\cite{LALsuite}. Posteriors obtained from simulated GR waveforms containing only the dominant ($\ell = 2, m = \pm 2$) modes observed by a single detector are in good agreement.}. From the posterior distribution $P(\bxi \, | \, d, H)$ of the full parameter set, we construct the posterior distribution $P(\Delta \blambda \, | \, d, H)$ of the set of parameters describing deviation from GR prediction, by marginalizing the posterior over all other parameters $\{\blambda, \btheta\}$. If the data is consistent with GR, we expect $P(\Delta \blambda \, | \, d, H)$ to be consistent with zero.

\paragraph{Simulations using GR waveforms:---}
We now demonstrate the power of the proposed test by making use of simulated GW observations from binary black holes, where the waveforms are modeled after the GR prediction. We employ the recent inspiral-merger-ringdown waveform model proposed by Mehta \etal~\cite{Mehta:2017jpq}, which provide accurate Fourier-domain models of the following spherical harmonic modes $\h_{\ell m}(f)$ of the expected GW signals from non-spinning binary black holes: $(\ell = 2, m = \pm2)$, $(\ell = 2, m=\pm1)$, $(\ell = 3, m=\pm3)$, $(\ell = 4, m = \pm4)$. (The other spherical harmonic modes that are neglected only introduce an inaccuracy (mismatch) of less than 1\% in the waveforms~\cite{Mehta:2017jpq}). GW observations are simulated by combining these signals with stationary Gaussian noise with power spectral density anticipated in Advanced LIGO's ``high-power, zero-detuning'' configuration~\cite{aLIGOZeroDetHighPower}, making use of Eqs.~(\ref{eq:spherical_harmonics}) and (\ref{eq:det_response}). We consider binaries with total mass $M := m_1 + m_2$ in the range $40 M_\odot$ -- 200 $M_\odot$ with mass ratio $q := m_2/m_1$ in the range 1/9 -- 1, with varying inclination angles $\iota$ (angle between the orbital angular momentum of the binary and the line of sight). 

\begin{figure}[tbh]
	\includegraphics*[width=3.5in]{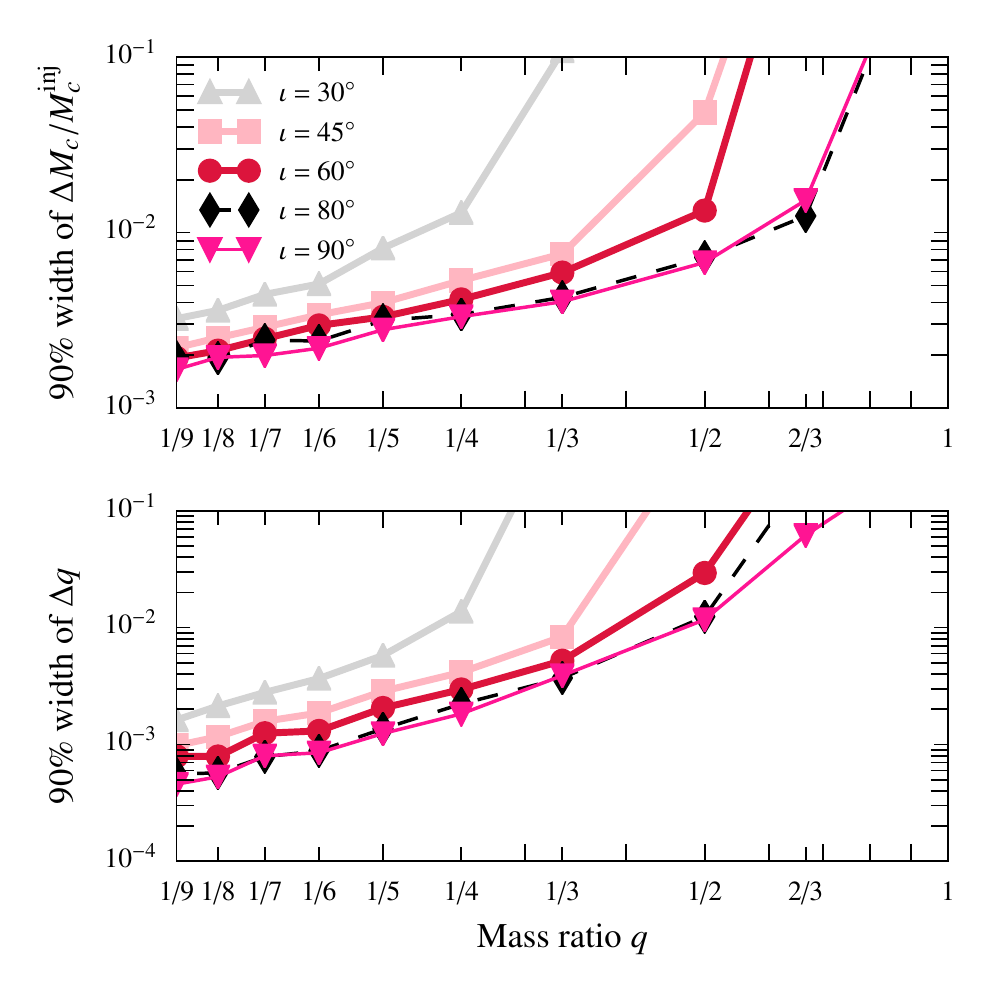}
	\caption{The figure shows the width of the 90$\%$ credible region of $\Delta M_c$ and $\Delta q$ for binaries with different mass ratios $q$ (horizontal axis) and inclination angles $\iota$ (legends). All binaries have a total mass $40M_{\odot}$. Best constraints are provided by binaries with high mass ratios and/or large inclination angles.}
	\label{fig:dMc_dq_posteriors_gr_vs_q}
\end{figure}

\begin{figure}[tbh]
	\includegraphics*[width=3.5in]{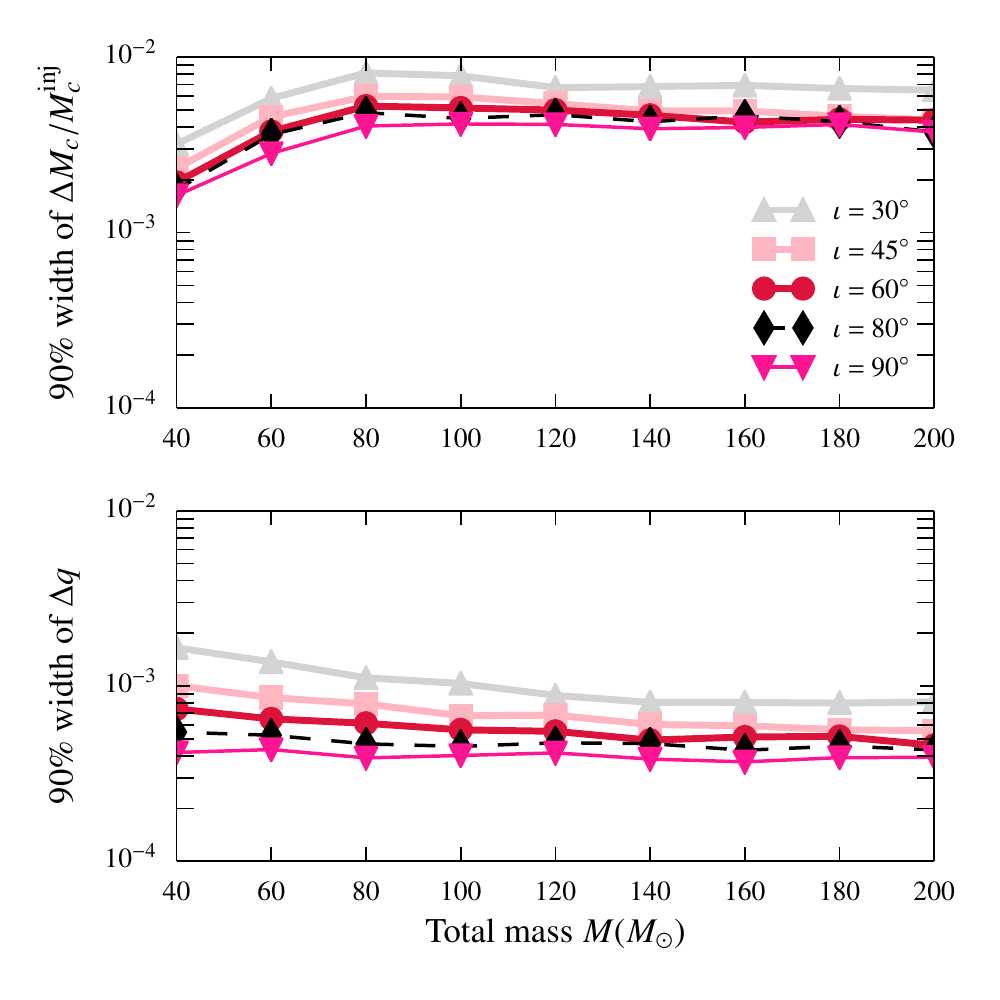}
	\caption{Same as Fig.~\ref{fig:dMc_dq_posteriors_gr_vs_q}, except that the horizontal axis reports the total mass $M$. All binaries correspond to a mass ratio $q = 1/9$.}
	\label{fig:dMc_dq_posteriors_gr_vs_M}
\end{figure}

We perform the test by introducing variations in the higher order modes: The higher-order modes $\hlm(f; \blambda+\Delta\blambda)$ are generated by introducing an extra parameter $\Delta\blambda$ while the quadrupole-modes $\h_{2\pm2}(f; \blambda)$ are generated by using the standard set of parameters $\blambda$ in GR. We have experimented with different choices for the deviation parameter $\Delta\blambda$: 
\begin{enumerate}
\item By introducing \emph{one} deviation parameter at a time. That is, $\Delta\blambda = {\Delta M_c}$ or $\Delta\blambda = {\Delta q}$. 
\item By introducing a concurrent deviation in \emph{two} parameters $\Delta \blambda = \{\Delta M_c, \Delta q\}$. 
\end{enumerate}
We show in Fig.~\ref{fig:contour_plots} the results of the tests performed by varying either one parameter or two parameters, for a binary with total mass $M = 80M_{\odot}$, mass ratio $q=1/9$, inclination angle $ {\iota}=60^{\circ} $ producing a signal-to-noise ratio  (SNR)  of 25 (SNR in higher modes is $\sim 10$). We see that the posterior probability density for the parameters $\Delta q$ and $\Delta M_c$ are consistent with zero as in GR. As expected, the width of the posterior is smaller when only one non-GR parameter is allowed to vary at a time. Figures~\ref{fig:dMc_dq_posteriors_gr_vs_q} and \ref{fig:dMc_dq_posteriors_gr_vs_M} show the 90\% credible regions of the posteriors of non-GR parameters for the case of binaries with different masses, mass ratios and inclination angles. In all cases, the SNR is set to {25}.  It is clear that binaries with large mass ratios ($q < 1/ 2$) and inclination angles ($\iota > 60 ^\circ $) will allow precision tests of the GR predictions, reaching statistical uncertainties of $< 10^{-2}$ for $\Delta M_c/M_c$ and $\Delta q$.   

\begin{figure}[tbh]
	\includegraphics*[width=3.5in]{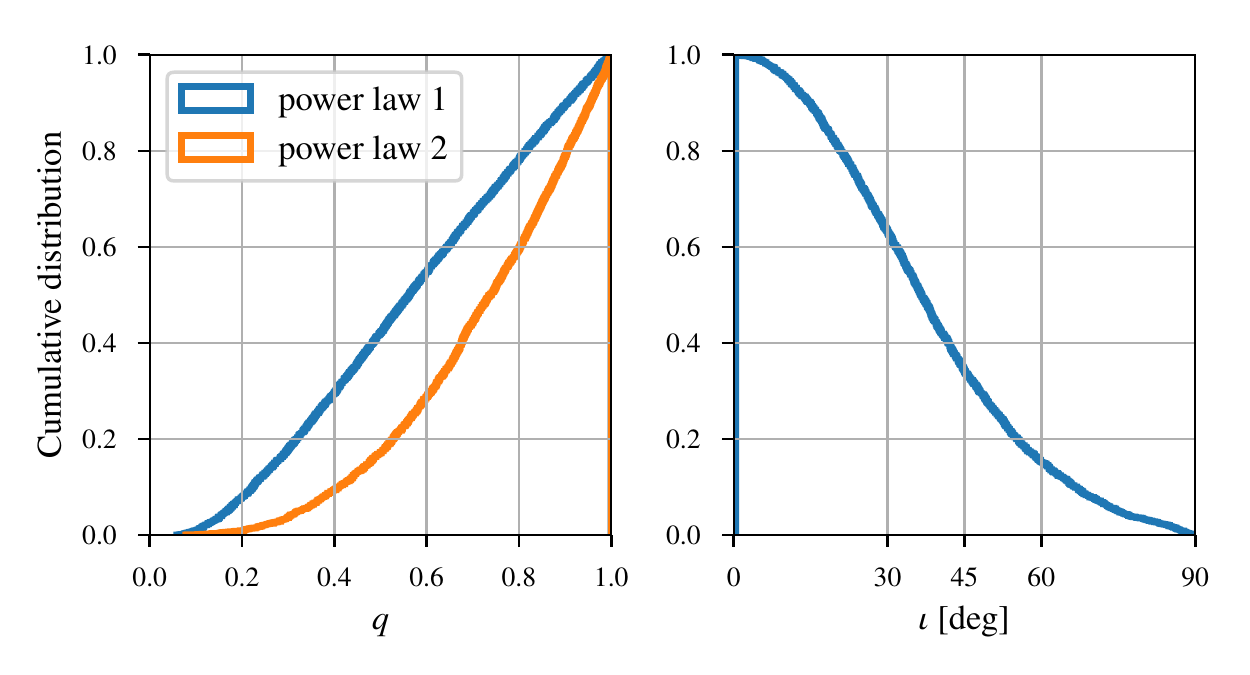}
	\caption{Projected cumulative distribution of the mass ratio $q$ (left) and inclination angle $\iota$ (right) of simulated binary black holes that are detectable by Advanced LIGO, based on our assumed component mass distribution. The two distributions in the left plot corresponds to two assumed distributions of the component masses (see text).}
	\label{fig:q_iota_distribution}
\end{figure}

\paragraph{Astrophysical prospects:---} 
Recent observations of GW signals from merging binaries of black holes~\cite{gw150914, gw151226, LSC_2016O1results, gw170104, gw170608, gw170814} and neutron stars~\cite{gw170817} by LIGO and Virgo have enabled the first tests of GR in the highly relativistic regime~\cite{LSC_2016grtests, LSC_2016O1results, gw170104, gw170608, gw170814}. However, the test proposed in this paper requires the observation of GW signals where the subdominant modes can be observed with appreciable SNR. These modes are excited predominantly for binaries with large mass ratios. Also, due to the radiation pattern, radiation from binaries with highly inclined orbits will contain appreciable contribution from subdominant modes. Hence binaries with large mass ratios ($q \lesssim 1/2$) and inclined orientations ($\iota \gtrsim 60^{\circ}$) are particularly suitable sources for performing the test described in this paper. Consequently, we do not expect the test to be effective for GW signals observed by LIGO and Virgo during their first two observational runs, for which mass ratios are less than 2 and inclinations are close to being face-on/face-off~\cite{LSC_2016O1results,gw170104,gw170608,gw170814}. The detection rate of binaries with large mass ratios depends on the astrophysical merger rate of such binaries, which is currently uncertain, while the detection rate of binaries with large inclination angle is related to the same with small inclination angles by a simple geometric factor. 

Here we investigate the  prospect of performing the proposed test on binary black hole events that Advanced LIGO and Virgo could observe over the next few years.  We simulate populations of binary black holes based on reasonable astrophysical assumptions, and examine the distributions of the mass ratio and inclination angle of detectable signals. In particular, we simulate binaries with two assumed distributions of component masses in the source-frame~\cite{Abbott:2016nhf}:
\begin{enumerate}
\item Masses following a power-law $p(m_{1,2}) = m_{1,2}^{-1}$   with $5 M_\odot \leq m_1, m_2  \leq 100 M_\odot$. 
\item Masses following a power-law $p(m_1) = m_1^{-2.35}$ on the mass of the larger black hole, with the smaller mass distributed uniformly in $q$ and with $5 M_\odot \leq m_1, m_2  \leq 100 M_\odot$. 
\end{enumerate}
In both cases, binaries are distributed uniformly in the sky with isotropic orientations. The distribution of the mergers in redshift is chosen according to the prescription given in~\cite{Dominik:2013tma}. The cosmological redshift on the GW signals can be absorbed by a rescaling of the masses $m_{1,2} (1+z)$ where $z$ is the redshift. From the simulated events, we compute the SNR expected  in Advanced LIGO and apply an SNR threshold for detection (the probability distributions are independent of the exact value of the SNR threshold). The cumulative distribution of the mass ratio $q$ and inclination angle $\iota$ of binaries crossing the detection threshold is plotted in Fig.~\ref{fig:q_iota_distribution}. It can be seen that $\sim 20 - 40\%$ of the detectable binaries will have a mass ratio greater than 2, out of which  $\sim 15\%$ will  be observed with inclination angle greater than $60^\circ$. Thus, only a few percent of the observed systems are likely to have large mass ratios ($q < 1/2$) and inclined orbits ($\iota > 60^\circ$). However, since Advanced LIGO and Virgo are expected to observe hundreds of binary black hole mergers over the next few years~\cite{Abbott:2016nhf}, we conclude that the proposed test could be performed when detectors reach their design sensitivity over the next few years, if not sooner.

\paragraph{Conclusions and future work:---} In this paper, we proposed a new method to test the consistency of an observed GW signal with a binary black hole system predicted by GR. The test relies on the fact that the multipolar structure of the radiated GW signal is uniquely determined in GR by the masses and spins of the black holes and no other parameters. Thus, if we estimate the parameters of the binary from different spherical harmonic modes of the observed signal independently, those estimates will have to be consistent with one another. Any inconsistency between the different estimates will point to a deviation from GR or to the non-black hole nature of the binary compact objects. We have used Bayesian parameter inference to identify potential deviations from GR predictions, using simulated GW signals.  We provided the first estimates of the expected precision of such tests that can be performed using GW observations of binary black holes anticipated by Advanced LIGO and Virgo in the next few years. 

The specific implementation of the test presented in this paper checks for the consistency of the masses (and spins, in the case of spinning binaries) estimated from the quadrupole/non-quadruple modes. If we have enough SNR to distinguish different modes, we can introduce deviation parameters for each mode (say, $\Delta M_c^{\ell m}$ and $\Delta q^{\ell m}$). This is analogous to checking the consistency of different quasi-normal mode frequencies, as the frequency evolution of the binary is determined by these intrinsic parameters. In addition, one could also check the consistency of the amplitudes of different modes, by introducing extra parameters describing deviations from the predicted amplitudes. While this would expand the scope of this test, in general, introducing more parameters would increase the statistical uncertainties, due to correlations between different parameters. 

In this paper we have assumed, for simplicity, that the companion black holes of the binary have negligible spins. Nevertheless, the method can be easily generalized to the case of binaries consisting of spinning black holes. We have also neglected the systematic errors due to inaccuracies in waveform modeling and detector calibration; these need to be understood before implementing the test on real observations. We leave these investigations to future work. 

\paragraph{Acknowledgments:---}
We thank Harald Pfeiffer, Bala Iyer, Gregorio Carullo, Vijay Varma, Juan Calderon Bustillo and Eric Thrane for useful comments on the manuscript, Chandra Kant Mishra for help with the numerical implementation of the waveform model used in this paper and M.~K.~Haris for the astrophysical simulations. This research was supported by the Indo-US Centre for the Exploration of Extreme Gravity funded by the Indo-US Science and Technology Forum (IUSSTF/JC-029/2016). PA's research was, in addition, supported by a Ramanujan Fellowship from the Science and Engineering Research Board, India and by the Max Planck Society through a Max Planck Partner Group at ICTS-TIFR and BSS's research was supported by NSF Grants AST-1716394 and AST-1708146. Computations were performed at the ICTS cluster Alice. 

\bibliographystyle{apsrev-nourl}
\bibliography{TGR_HM}

\end{document}